\documentclass[aps,amsmath, prb,twocolumn,superscriptaddress]{revtex4-2}
\usepackage{lineno}
\usepackage{graphicx}
\usepackage{dcolumn}
\usepackage{bm}
\usepackage{lipsum}
\usepackage{color}
\usepackage{amsfonts,eucal,mathrsfs,amssymb}
\DeclareMathAlphabet{\mathscrbf}{OMS}{mdugm}{b}{n}

\usepackage[table]{xcolor}
\usepackage{multirow}
\usepackage[breaklinks=true,unicode=true,urlcolor = blue,colorlinks = true,citecolor = blue,linkcolor = blue]{hyperref}

\renewcommand{\vec}[1]{\bm{#1}}
\newcommand{\dd}{\mathrm{d}}

\begin{document}

\preprint{}
\title{Curvature induced magnetization of altermagnetic films}

\author{Kostiantyn V. Yershov}
\affiliation{Leibniz-Institut f\"{u}r Festk\"{o}rper- und Werkstoffforschung, Helmholtzstraße 20, D-01069 Dresden, Germany}
\affiliation{Bogolyubov Institute for Theoretical Physics of the National Academy of Sciences of Ukraine, 03143 Kyiv, Ukraine}

\author{Olena Gomonay}
\affiliation{Institut f\"{u}r Physik, Johannes Gutenberg-Universit\"{a}t Mainz, Staudingerweg 7, D-55099 Mainz, Germany}

\author{Jairo Sinova}
\affiliation{Institut f\"{u}r Physik, Johannes Gutenberg-Universit\"{a}t Mainz, Staudingerweg 7, D-55099 Mainz, Germany}
\affiliation{Department of Physics, Texas A\&M University, College Station, Texas 77843-4242, USA}

\author{Jeroen van den Brink}
\affiliation{Leibniz-Institut f\"{u}r Festk\"{o}rper- und Werkstoffforschung, Helmholtzstraße 20, D-01069 Dresden, Germany}

\author{Volodymyr P. Kravchuk}
\affiliation{Leibniz-Institut f\"{u}r Festk\"{o}rper- und Werkstoffforschung, Helmholtzstraße 20, D-01069 Dresden, Germany}
\affiliation{Bogolyubov Institute for Theoretical Physics of the National Academy of Sciences of Ukraine, 03143 Kyiv, Ukraine}

\begin{abstract}
We consider a thin film of $d$-wave altermagnet bent in a stretching-free manner and demonstrate that gradients of the film curvature induce a local magnetization which is approximately tangential to the film. The magnetization amplitude directly reflects the altermagnetic symmetry and depends on the direction of bending. It is maximal for the bending along directions of the maximal altermagnetic splitting of the magnon bands. A periodically bent film of sinusoidal shape possesses a total magnetic moment per period $\propto\mathscr{A}^2q^4$ where $\mathscr{A}$ and $q$ are the bending amplitude and wave vector, respectively. The total magnetic moment is perpendicular to the plane of the unbent film and its direction (up or down) is determined by the bending direction. A film roll up to a nanotube possesses a toroidal moment directed along the tube $\propto \delta_r/r^2$ per one coil, where $r$ and $\delta_r$ are the coil radius and the pitch between coils. 
All these analytical predictions agree with numerical spin-lattice simulations.
\end{abstract}

\maketitle

\section{Introduction}
Altermagnetism is a recently emergent and rapidly growing domain in the physics of magnetically ordered solids~\cite{Smejkal22a}. Being collinear-compensated magnets, altermagnets differ from conventional antiferromagnets by a more complex symmetry transformation connecting two sub-lattices. Due to the specific local surrounding of the magnetic atoms, the symmetry transformation involves also the rotation operation in addition to the translation and time reversal~\cite{Smejkal20,Smejkal22}. The latter lifts the degeneracy of the nonrelativistic electron~\cite{Smejkal22,Smejkal20,Ahn19,Yuan20,Ma21} and magnon~\cite{Naka19,Smejkal23,Gohlke23,Gomonay24} bands such that the band splitting possesses $d$/$g$/$i$-wave symmetry.

Recently a phenomenological model of $d$-wave altermagnets was proposed~\cite{Gomonay24}, and a number of new properties of noncollinear magnetic textures were predicted. In particular, it was shown that a static domain wall possesses a locally distributed magnetization even in zero magnetic field~\cite{Gomonay24}. The effect essentially depends on the domain wall orientation, and it is maximal when gradients of the N{\'e}el order parameter are in the directions of the maximal splitting of the magnon bands. This phenomenon can be intuitively understood in terms of the different effective exchange stiffnesses in the different sublattices. As a consequence, the magnetic compensation of the noncollinear texture is incomplete due to the different sub-lattice length scales; the latter gives rise to the non-compesated magnetic moments.

Here we generalize the previously developed phenomenology of the $d$-wave altermagnetic films~\cite{Gomonay24} for the case of a curvilinear film bent in a stretching-free manner. We predict a curvature-induced mechanism of the magnetization generation and demonstrate that the bending of an altermagnetic film can induce magnetization also for the case of a colinear magnetic ordering. The maximal effect is expected for the bends in the directions of the maximal splitting of the magnon bands. It is known~\cite{Gaididei14,Sheka15,Ortix23,Yershov22,Kravchuk18a,Pylypovskyi18a,Yershov15b,Korniienko19b} that in the curvilinear films and wires, the competition between the isotropic exchange interaction and uniaxial magnetocrystalline anisotropy, whose orientation follows the magnet geometry~\footnote{E.g. the anisotropy easy-axis is normal to the film or tangential to the wire.}, can lead to deviation of the magnetization order parameter from the equilibrium easy-axial (easy-planar) orientation. This effect can be understood as an action of some effective magnetic field~\cite{Gaididei14,Sheka15}, which in torsion-free geometries is determined by the curvature gradients. As a result, an altermagnetic film bent with nonzero curvature gradients possesses a distribution of the N{\'e}el order parameter which is nonuniform in the local curvilinear frame of reference. The non-uniformity of the order parameter leads to the generation of magnetization via the previously established mechanism~\cite{Gomonay24}. In the limit of small curvature gradients, the generated magnetization is linear in the curvature gradients and approximately tangential to the film. The nonzero net magnetization can be generated for the case of a periodic deformation with alternating of the curvature gradients sign. The latter situation is realized for the case of the sinusoidally rippled film which we consider below as an example. Interestingly, the intuitively expected inverse effect of the altermagnetic film rippling by the applied magnetic field does not take place.

\section{Model and notations}

\begin{figure}
	\includegraphics[width=\columnwidth]{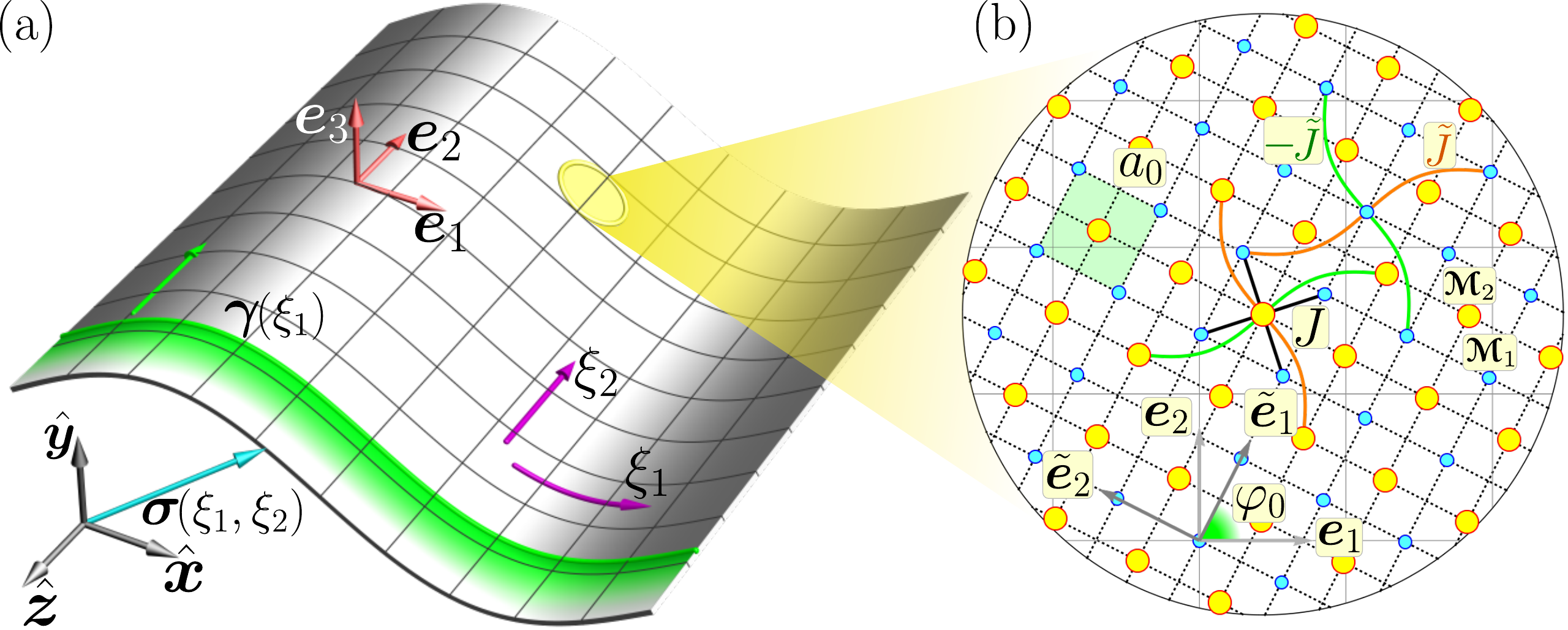}
	\caption{A double-layer altermagnetic film is deformed in form of a generalized cylinder surface $\vec{\sigma}$ which is formed by the parallel translation of the directrix $\vec{\gamma}\in x0y$ along generatrix $-\hat{\vec{z}}$ (green arrow). On the surface, we introduce curvilinear orthogonal coordinates $\xi_1$ and $\xi_2$ such that $\xi_1$ is the arclength of $\vec{\gamma}$ and $\xi_2=-z$, see panel (a). The orientation of the crystalographic axes (directions $\tilde{\vec{e}}_\alpha$) with respect to the basis $\vec{e}_\alpha$ is controlled by the angle $\varphi_0$, see panel (b). }\label{fig:bending}
\end{figure}

\begin{figure*}
	\includegraphics[width=\textwidth]{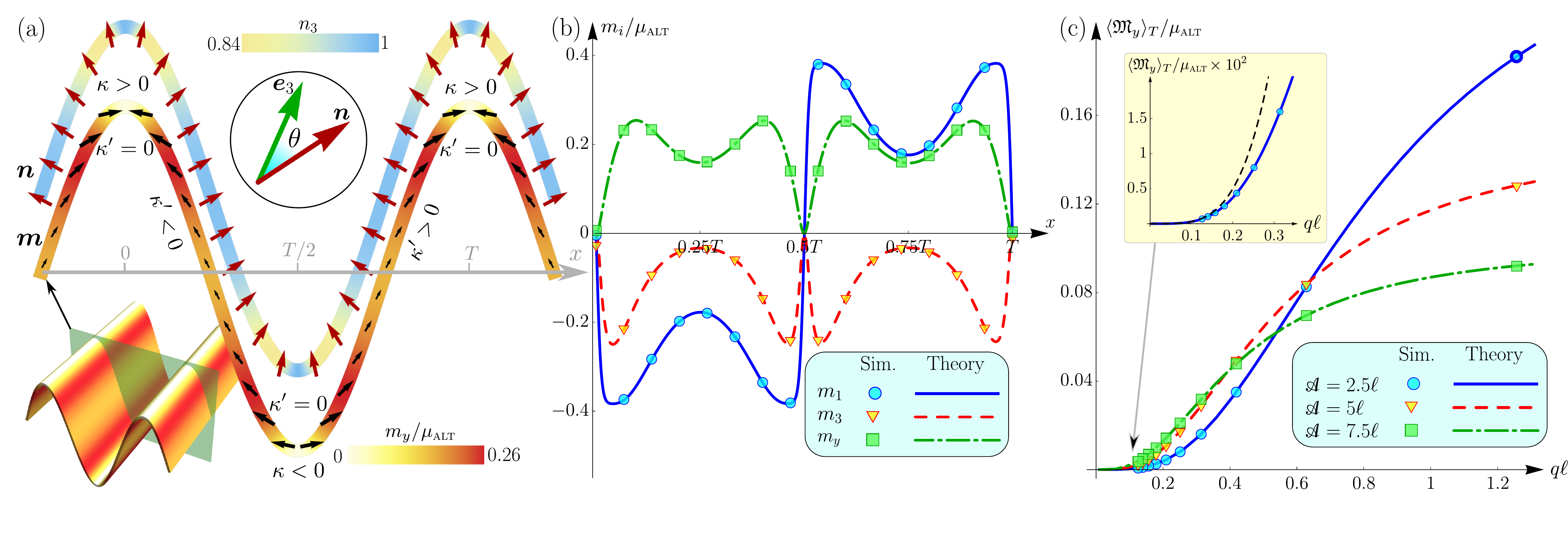}
	\caption{The curvature induced magnetization of a sine-shaped film. Panel (a) shows the equilibrium distributions of vectors $\vec{n}$ (top) and $\vec{m}$ (bottom) obtained from spin-lattice simulations which reproduce the dependencies determined by equations \eqref{eq:theta} and \eqref{eq:m}, respectively. The color schemes represent the normal component $n_3=\vec{n}\cdot\vec{e}_3$ of the N{\'e}el vector and the magnetization component $m_y$ which is normal to the undeformed film. On panel (b), we compare different magnetization components obtained theoretically by means of formula \eqref{eq:m} and by means of spin-lattice simulations (see Appendix~\ref{app:simuls}). Panel (c) shows the total magnetic moment per period for various wave vectors $q=2\pi/T$ and amplitudes of the sinusoidal deformation. The highlighted marker corresponds to the parameters of the panels (a) and (b). The inset demonstrates asymptotics \eqref{eq:mT} (dashed line). }\label{fig:m-sin}
\end{figure*}

Here we consider a thin film of a $d$-wave altermagnet with the crystal structure of rutile, e.g.  MnF$_2$, CoF$_2$, RuO$_2$. The film is grown in plane (001) and bent in the form of a generalized cylindrical surface. The latter can be treated as 2D locus swept by a planar curve $\vec{\gamma}\in x0y$ by its parallel translation is $z$-direction, see Fig.~\ref{fig:bending}(a). On the surface, we introduce a local tangential basis $\vec{e}_\alpha$ with $\alpha=1,2$ such that $\vec{e}_1$ is a unit vector tangential to $\vec{\gamma}$, and $\vec{e}_2=-\hat{\vec{z}}$. The corresponding curvilinear coordinates $\xi_\alpha$ are such that $\xi_1$ is the arclength of $\vec{\gamma}$ and $\xi_2=-z$. Vectors $\vec{e}_1$ and $\vec{e}_2$ are orthogonal by the construction, thus $\vec{e}_3=\vec{e}_1\times\vec{e}_2$ is a unit normal. The considered cylindrical surface possesses zero Gau{\ss}ian curvature and therefore it is free of stretching. The latter means that a two-dimensional discrete lattice can be fit to the curvilinear surface with preserving distances and angles between the neighboring atoms. As a model, we consider a magnet with two square sublattices bounded by antiferromagnetic exchange $J$, see Fig.~\ref{fig:bending}(b). The additional next to the nearest neighbors exchange interactions $\pm\tilde{J}$ with symmetries shown in Fig.~\ref{fig:bending}(b) makes the considered model altermagnetic~\cite{Gomonay24}. Additionally, we take into account the uniaxial anisotropy with the easy-axis oriented along the normal $\vec{e}_3$. It was shown~\cite{Gomonay24} that the proposed model captures properties of a double-layer film of RuO$_2$. 
In the following, we consider a general case when the bending direction $\xi_1$ makes an arbitrary angle $\varphi_0$ with the direction $\tilde{\vec{e}}_1$ of the crystalographic axis [100], see Fig.~\ref{fig:bending}(b).

In the following we utilize the continuous description in which magnetization of each sublattice is represented by continuous vector function $\vec{M}_\alpha(\xi_1,\xi_2)$ of the constant absolute value $|\vec{M}_\alpha|=M_s$. Here $M_s=\mu_s/(a_0^2c_0)$ is the saturation magnetization of one sublattice with $\mu_s$ being magnetic moment of one lattice site (e.g. Ru), and $a_0$ and $c_0$ are the lattice constants in the directions tangential and perpendicular to the film, respectively. It is instructive to introduce the dimensionless vector of magnetization $\vec{m}=(\vec{M}_1+\vec{M}_2)/(2M_s)$ and N{\'e}el vector $\vec{n}=(\vec{M}_1-\vec{M}_2)/(2M_s)$.

In the limit $|\vec{m}|\ll1$, dynamics of the N{\'e}el vector is determined by the action $\mathcal{S}=h\int\dd t\iint\dd\xi_1\dd\xi_2\mathscr{L}$ with Lagrangian
\begin{equation}\label{eq:L}
	\begin{split}
		&\mathscr{L}=\frac{M_s}{\gamma_0^2B_{ex}}\left[\dot{\vec{n}}+\vec{n}\times\gamma_0\vec{B}\right]^2-A_{\textsc{afm}}\,\partial_\alpha\vec{n}\cdot\partial_\alpha\vec{n}\\
		&-K[\vec{n}\times\vec{e}_3]^2+\frac{2A_{\textsc{alt}}}{\gamma_0B_{ex}}\left[(\dot{\vec{n}}+\vec{n}\times\gamma_0\vec{B})\times\vec{n}\right]\cdot\hat{\mathfrak{D}}\vec{n}.
	\end{split}	
\end{equation}
Here $h$ is thickness of the film, $\gamma_0$ is gyromagnetic ratio, $B_{ex}=8J/\mu_s$ is exchange field, $\vec{B}$ is the external applied magnetic field, $K>0$ is constant of the easy-normal anisotropy, $A_{\textsc{afm}}=J/(2c_0)$ is the antiferromagnetic stiffness, and constant $A_{\textsc{alt}}=4\tilde{J}/c_0$ determines strength of the altermagnetic effects. We introduce differential operator $\hat{\mathfrak{D}}=\partial^2_{\tilde{\xi}_1\tilde{\xi}_2}=\cos2\varphi_0\partial^2_{12}-\sin\varphi_0\cos\varphi_0(\partial^2_{11}-\partial^2_{22})$ where $\tilde{\xi}_\alpha$ are the coordinates in directions of the crystalographic axes $\tilde{\vec{e}}_\alpha$, and we denoted $\partial_\alpha=\partial_{\xi_\alpha}$ for the sake of simplicity. Here we also took into account that the metric is Euclidean in the reference frame $(\xi_1,\xi_2)$. The N{\'e}el vector obtained from \eqref{eq:L} determines magnetization
\begin{equation}\label{eq:m}
	\vec{m}=\frac{1}{\gamma_0B_{ex}}\left[\dot{\vec{n}}+\vec{n}\times\gamma_0\vec{B}\right]\times\vec{n}+\eta\,\vec{n}\times\hat{\mathfrak{D}}\vec{n}\times\vec{n},
\end{equation}
where $\eta=A_{\textsc{alt}}/(M_sB_{ex})=\tilde{J}a_0^2/(2J)$. For the derivation of \eqref{eq:L} and \eqref{eq:m} see Appendix~\ref{app:models}. Note that due to the last term in \eqref{eq:m}, the magnetization can appear even in a static case without external field. This causes a purely altermagnetic effect, and the corresponding magnetization induced by domain walls (DW).

\section{The curvature induced magnetization}\label{sec:induced_m}

Let us first consider a static case without applied magnetic field. In this case, according to \eqref{eq:L}, the equilibrium solution for $\vec{n}$ is not affected by the altermagnetism and one can show~\cite{Ortix23} (see also Appendix~\ref{app:geom}) that it has the following properties: $\vec{n}\in x0y$, and $\vec{n}=\vec{n}(\xi_1)$, e.g. the N{\'e}el vector lies in the same plane as directrix $\vec{\gamma}$ and depends only on its arclength. Thus, the orientation of the N{\'e}el vector is represented by the only one angle $\theta(\xi_1)$ in the way $\vec{n}=\sin\theta\vec{e}_1+\cos\theta\vec{e}_3$. Angle $\theta$ is determined by the differential equation
\begin{equation}\label{eq:theta}
	\theta''-\frac{1}{\ell^2}\sin\theta\cos\theta=-\kappa',
\end{equation}
where prime denotes derivative with respect to $\xi_1$, and $\ell=\sqrt{A_{\textsc{afm}}/K}$ is typical length-scale of the system, and $\kappa=[\vec{\gamma}''\times\vec{\gamma}']\cdot\hat{\vec{z}}$ is curvature of the directrix which is also the mean curvature of the surface. This definition results in positive and negative curvature sign for the ``tops'' and ``valleys'' of the surface, respectively. Note that due to the curvature gradients, N{\'e}el vector deviates from the normal direction. This effect is well known for curvilinear films~\cite{Yershov22,Kravchuk18a,Pylypovskyi18a,Ortix23} and wires~\cite{Yershov15,Sheka15,Korniienko19b} with perpendicular anisotropy.

 Since $\vec{n}$ depends on $\xi_1$ only, according to \eqref{eq:m} and to the definition of operator $\hat{\mathfrak{D}}$, the maximal magnetization amplitude is expected for $\varphi_0=\pi/4+\nu\pi/2$ with $\nu\in\mathbb{Z}$, i.e. when the bending is made along diagonals of crystal lattice. In what follows, we focus on the case $\varphi_0=-\pi/4$ and therefore the magnetization \eqref{eq:m} is
\begin{subequations}\label{eq:m-smpl}
	\begin{align}
	\label{eq:m-smpl-n}\vec{m}=&\frac{\eta}{2}\vec{n}\times\vec{n}''\times\vec{n}\\
\label{eq:m-smpl-theta}	=&\mu_{\textsc{alt}}\sin\theta\cos\theta\left(\vec{e}_1\cos\theta-\vec{e}_3\sin\theta\right).
	\end{align}
\end{subequations}
Deriving the second expression, we took into account the coordinate dependence of the local basis $\vec{e}_i$ and utilized Eq.~\eqref{eq:theta}. The dimensionless parameter $\mu_{\textsc{alt}}=\frac12 A_{\textsc{alt}}B_{an}/(A_{\textsc{afm}}B_{ex})$ is the typical scale for the curvature induced magnetization in the system. Here $B_{an}=K/M_s$ is the anisotropy field.

According to \eqref{eq:m-smpl-theta} the deviation of the Ne{\'e}el vector from the strictly normal direction ($\theta=0,\,\pi$) is necessary for the magnetic moment generation. For example, surface of a circular cylinder of radius $R$ possesses constant curvature $\kappa=1/R$. Since the right hand side part of Eq.~\eqref{eq:theta} vanishes in this case, the equilibrium state is simply $\theta=0,\,\pi$ meaning $\vec{n}=\pm\vec{e}_3$. And according to \eqref{eq:m-smpl-theta} the curvature induced magnetization does not appear. Next, we consider two specific cases: (i) the wave-shaped periodically deformed film, and (ii) a film rolled up into a nanotube. 

\emph{Wave-shaped film.} This geometry we realize by choosing the directrix in the form $\vec{\gamma}=x\hat{\vec{x}}+f(x)\hat{\vec{y}}$ with $f=\mathscr{A}\cos(qx)$. Formation of films of such a film can be experimentally accessible via thermal scanning probe lithography~\cite{Howell20,Lassaline23}. Computing the curvature $\kappa=-f''_{xx}/(1+f'^2_x)^{3/2}$, we use Eq.~\eqref{eq:theta}\footnote{Naturally, we take into account the relation between coordinates $\dd\xi_1=\sqrt{1+f'^2_x}\dd x$.} in order to find distribution of the N{\'e}el vector along the surface, see Fig.~\ref{fig:m-sin}(a).

The competition between exchange stiffness and the anisotropy whose easy-axis follows the film geometry results in the regions (yellow) where the N{\'e}el vector deviates from the normal direction. This is consistent with the recent results obtained for ferromagnets~\cite{Ortix23}. According to Eq.~\eqref{eq:m-smpl-theta}, such a deviation leads to generation of magnetization in these regions, see the bottom part of Fig~\ref{fig:m-sin}(a). The induced magnetization is approximately tangential to the surface. Indeed, in the limit of small curvature gradients $|\kappa'|\ell^2\ll1$, one obtains from \eqref{eq:theta} the approximate solution $\theta\approx\ell^2\kappa'$ and consequently, Eq.~\eqref{eq:m-smpl-theta} results in $\vec{m}\approx\mu_{\textsc{alt}}\ell^2\kappa'\vec{e}_1$. Magnetization vanishes in the points of the curvature extremum ($\kappa'=0$) which corresponds to the highest and lowest points of the film profile. Between the extremum point, the curvatre gradient $\kappa'$ flips sign, which causes the flip of sign of $\theta$ and the tangential component $m_1$, see Fig.~\ref{fig:m-sin}(b). As a result, the transversal component $m_y$ does not change the sign giving rise to the total transversal magnetization per period $\langle \vec{\mathfrak{M}}\rangle_T=V^{-1}_{T}\int_{V_T}\vec{m}\,\dd\vec{r}=\hat{\vec{y}}\langle\mathfrak{M}_y\rangle_T$, with $V_T$ being the volume of one period of the film,  see  Fig.~\ref{fig:m-sin}(c). Note the perfect agreement between the theoretical predictions (lines) based on continuum Eqs.~\eqref{eq:theta}, \eqref{eq:m} and discrete spin-lattice simulations (markers), for details see Appendix~\ref{app:simuls}. In simulations, we model the overdamped dynamics of a system of approximately $10^5$ spins by solving numerically a set of coupled Landau-Lifshits-Gilbert equations, the number of which is equal to the number of spins, for details see Appendix~\ref{app:simuls}.

\begin{figure}
	\includegraphics[width=0.75\columnwidth]{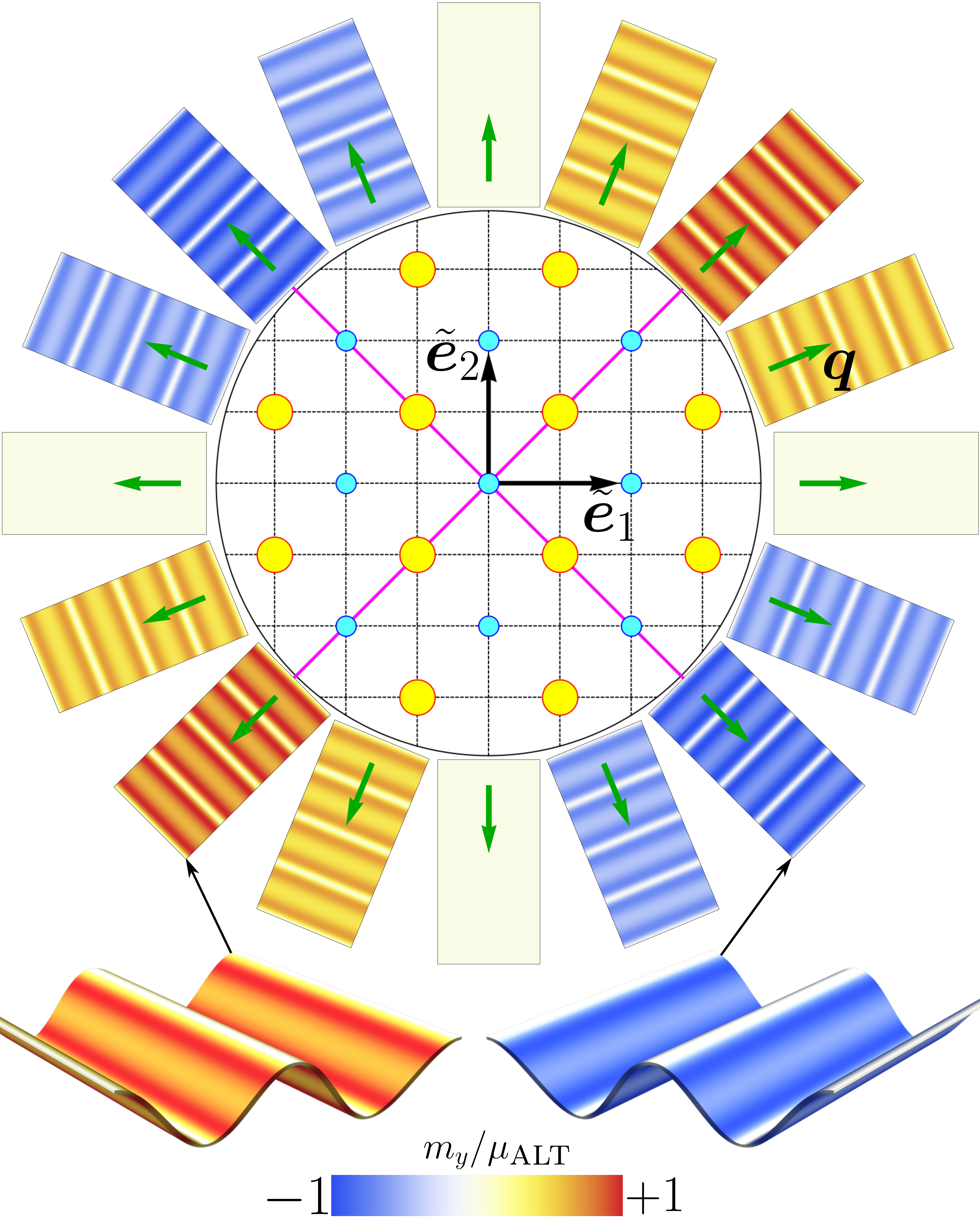}
	\caption{Distribution of the curvature-induced transversal component $m_y$ of magnetization over the sinusoidally deformed stripes (view from above) for various orientations of the deformation vector $\vec{q}$ with respect to the crystallographic directions. The magenta lines show the direction of $\vec{q}$ which corresponds to the maximum magnetization amplitude.}\label{fig:m-q}
\end{figure}

\begin{figure}
	\centering
	\includegraphics[width=\columnwidth]{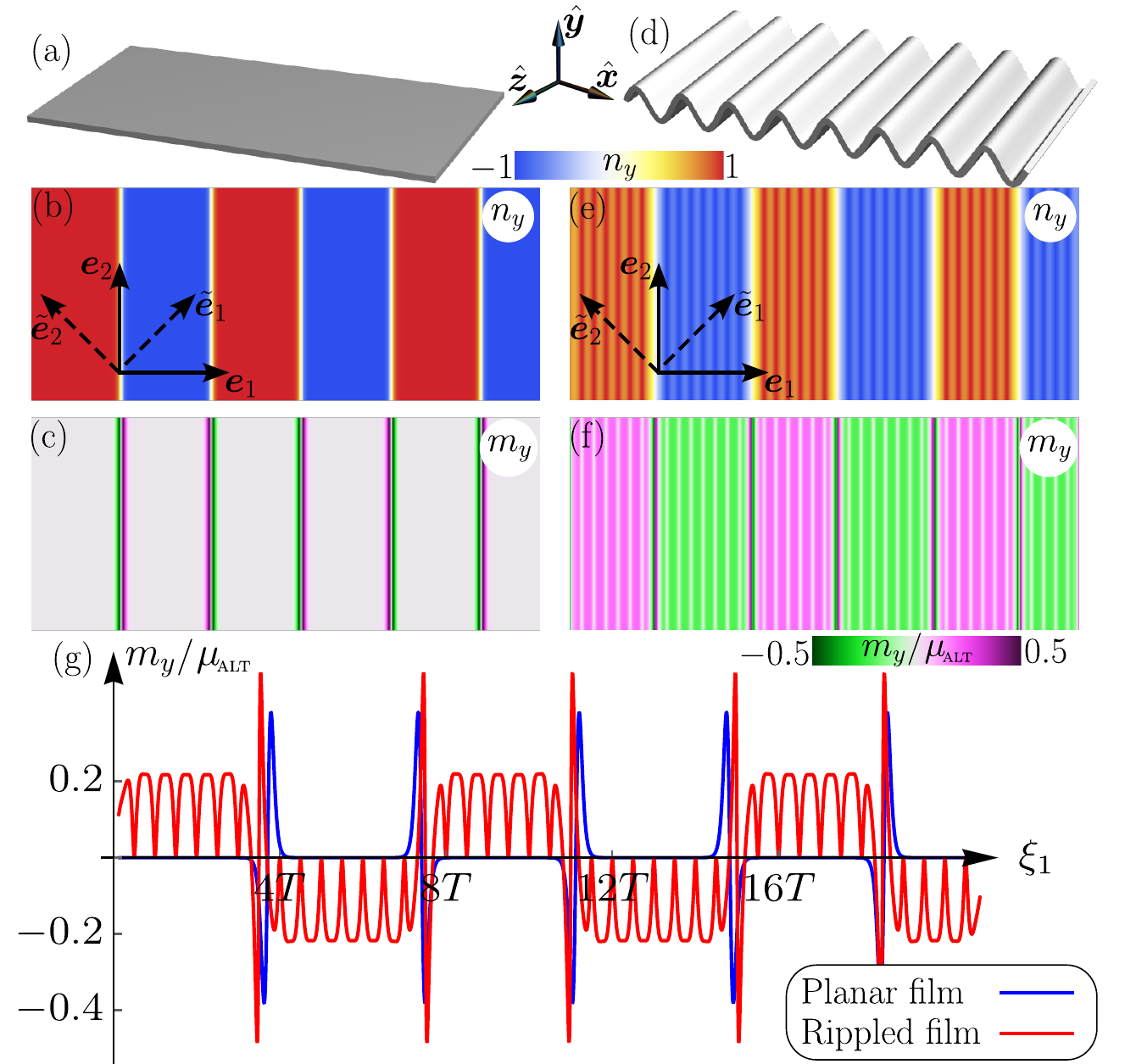}
	\caption{Utilization of the film rippling to manifest the domain pattern in the magnetization channel. Panels (a,b,c) and (d,e,f) correspond to the planar and sinusoidal rippled films, respectively. (a,d) -- Schematically shown geometries. Panels (b,e) and (c,f) show the domain patterns in terms of N{\'e}el vector and magnetization, respectively. The magnetization distributions along $\vec{e}_1$, the direction perpendicular to DWs, are compared in panel (g) for the planar and rippled cases. In both cases, the domain pattern was relaxed using the spin-lattice simulations. Parameters of the rippling are $T=5\ell$, $\mathscr{A}=1.5\ell$, for more details see Appendix~\ref{app:simuls}.}
	\label{fig:DWs}
\end{figure}

In the limit of small curvature gradients, we approximate $\kappa'\approx-f'''_{xxx}$ and estimate the asymptotics 
\begin{equation}\label{eq:mT}
	\langle \mathfrak{M}_y\rangle_T\approx\frac{\mu_{\textsc{alt}}}{2}\mathscr{A}^2q^4\ell^2,
\end{equation}
see the dashed line in Fig.~\ref{fig:m-sin}(c). Here we consider the case when direction of the deformation wave vector $\vec{q}$ corresponds to maximal curvature induced magnetization. The cases which correspond to all possible orientations of $\vec{q}$ are summarized in Fig.~\ref{fig:m-q}.

The effect of curvature-induced magnetization opens up new possibilities in the investigation of altermagnets. Namely, due to the geometric rippling of the film any domain structure of the antiferromagnetic compensated ordering will become visible in the magnetization channel, as is demonstrated in Fig.~\ref{fig:DWs}(f). This is because the geometry-induced part of the magnetization~\eqref{eq:m} flips sign under the transformation $\vec{n}\to-\vec{n}$. Therefore, the domains corresponding to different ground states acquire the curvature-induced magnetization of opposite signs. To demonstrate this feature we use the spin-lattice simulations and relax the same domain patterns for the planar and rippled films. The results are compared in Fig.~\ref{fig:DWs}. For a planar film, the magnetization is induced by the gradients of $\vec{n}$ only and it is strongly localized on the domain walls (DW)~\cite{Gomonay24a}, see blue line in Fig.~\ref{fig:DWs}(g). Moreover, this magnetization has opposite signs from the different sides of the DW, resulting in an almost vanishing stray field. The latter complicates the DW observation. For a rippled film, entire domains acquire a magnetization, see the red line in Fig.~\ref{fig:DWs}(g). This allows the well-developed set of experimental methods to observe ferromagnetic domains~\cite{Hubert98} and their dynamics to be utilized as well to resolve altermagnetic domains via its characteristic magnetization patterns, e.g. as shown in Fig.~\ref{fig:DWs}(f).

We have demonstrated that the periodic rippling of the altermagnetic film leads to generation of the averaged magnetization $\langle\vec{m}\rangle_T$ in the transversal (along $\hat{\vec{y}}$) direction. This raises a question about the possibility of an inverse effect, namely the inducing a rippling of the film by applying an external magnetic field $\vec{B}=B\hat{\vec{y}}$. The analysis of the self-consistent problem in applied magnetic field, which includes both elastic and magnetic degrees of freedom, shows the stability of the planar film solution under the condition $(A_{\textsc{alt}}/A_{\textsc{afm}})(B/B_{ex})<1$, for details see Appendix~\ref{app:elast}. Thus, the inverse effects of the field-induced film deformation is not expected for the physically realistic conditions.

\begin{figure}[t]
	\includegraphics[width=\columnwidth]{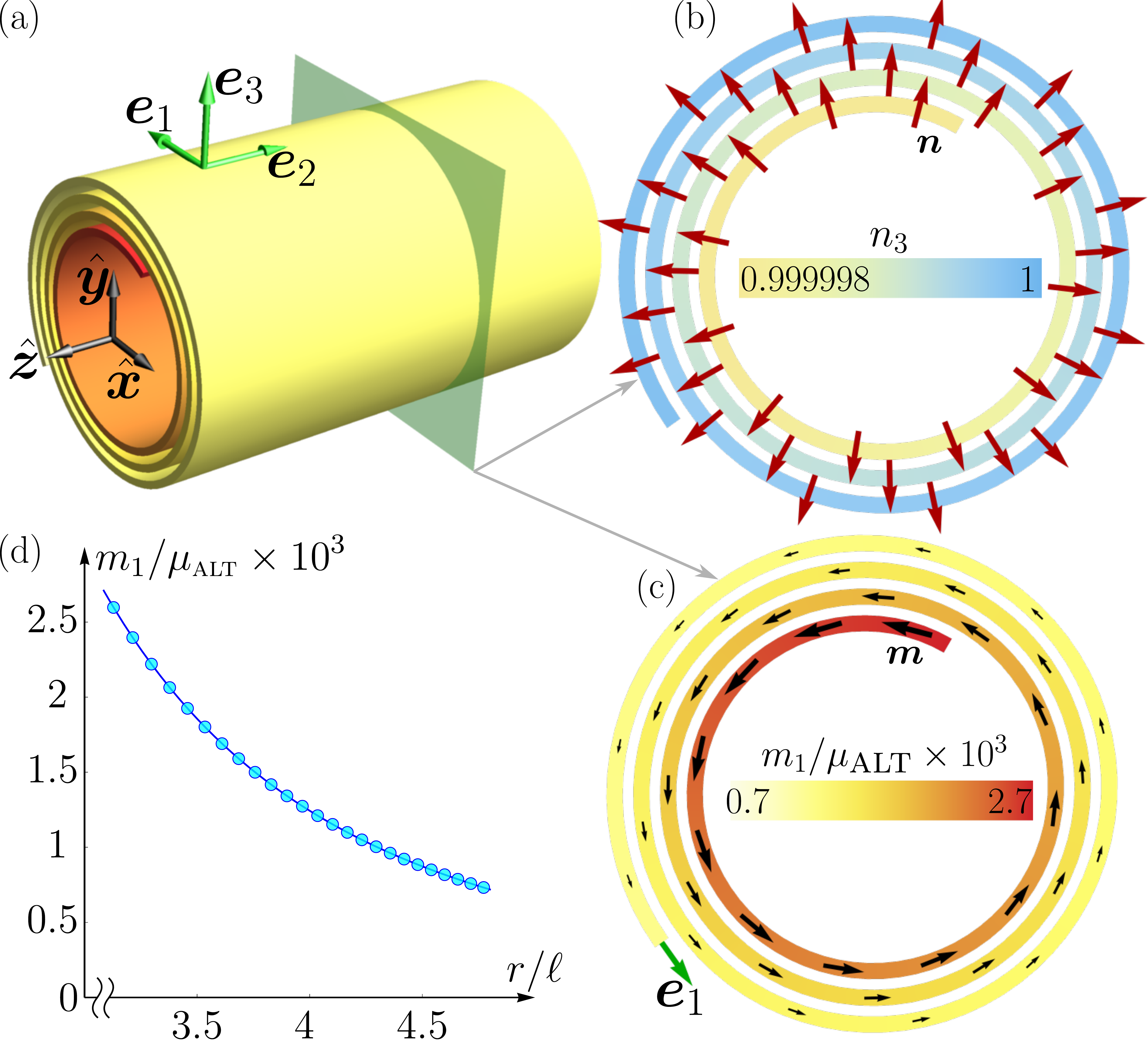}
	\caption{(a) The curvature-induced magnetization of the film rolled up in form of an Arhimedean spiral with pitch $\delta_r=0.5\ell$. Panels (b)-(c) show the equilibrium distribution of vectors $\vec{n}$ and $\vec{m}$, respectively, obtained by means of numerical simulations. The color schemes are the same as in Fig.~\ref{fig:m-sin}. Panel (d) shows a comparison of the tangential magnetization obtained from spin-lattice simulations (markers) and the analytical approximation $m_1\approx\mu_{\textsc{alt}}\ell^2\delta_r/(2\pi r^3)$ (line).}\label{fig:m-aSpiral}
\end{figure}

\emph{Rolled up nanotube.} Let us now consider a case when the altermagnetic film is rolled up in a nanotube. Experimental realizations of such nanotubes were recently reported~\cite{Schmidt01a,Grimm12}. We model the roll-up tube by a cylindrical Archimedean spiral with directrix $\vec{\gamma}=r(\chi)(\hat{\vec{x}}\cos\chi+\mathcal{C}\hat{\vec{y}}\sin\chi)$ where $r=\delta_r\chi/(2\pi)$ and $\chi>0$, see Fig.~\ref{fig:m-aSpiral}(a). Here $\delta_r>0$ and $\mathcal{C}=\pm1$ are the spiral pitch and chirality, respectively.  The Archimedean spiral possesses a nonvanishing curvature gradient giving rise to the deviation of the N{\'e}el vector from the normal direction and to the curvature-induced magnetization shown in Fig.~\ref{fig:m-aSpiral}(b) and (c), respectively. In the limit $r\gg\delta_r$, the curvature gradient is $\kappa'\approx\mathcal{C}\delta_r/(2\pi r^3)$. Therefore, in this limit, we expect the curvature induced magnetization $\vec{m}\approx\mu_{\textsc{alt}}\ell^2\delta_r\vec{e}_1/(2\pi r^3)$ where the tangential vector $\vec{e}_1$ is directed towards the unfolding of the spiral, and we assumed that N{\'e}el vector is directed outward the spiral, see Fig.~\ref{fig:m-aSpiral}(b,c). The proposed approximation for $\vec{m}$ perfectly agrees with the magnetization obtained in the spin-lattice simulations, see Fig.~\ref{fig:m-aSpiral}(d). 

The tangential magnetization gives rise to the toroidal moment \cite{Ederer07} per one spiral coil
\begin{equation}\label{eq:T}
	\langle\vec{T}\rangle_C=\frac{1}{2V_\textsc{c}}\int_{V_\textsc{c}}[\vec{r}\times\vec{m}]\dd\vec{r}\approx\frac{\mu_{\textsc{alt}}\ell^2\delta_r\mathcal{C}\hat{\vec{z}}}{4\pi r^2},
\end{equation}
where $V_\textsc{c}$ is the spiral coil volume. The presence of the toroidal moment allows the antisymmetric magnetoelectric coupling~\cite{Spaldin08}.

\section{Conclusions}
The stretching-free bending of a thin altermangnetic film with nonzero curvature gradients induces local magnetization that is approximately tangential to the film. The magnitude of the curvature-induced magnetization is determined by the curvature gradient and the direction of bending relative to the crystallographic axes. The maximal effect is achieved when bending is in the direction of maximum altermagnetic splitting in the magnon spectrum. The effect of the curvature-induced magnetization is expected for also $g$- and $i$-wave altermagnets, however with higher sensitivity concerning the bending direction.

Due to the effect of the curvature-induced magnetization, the specific film geometries can possess macroscopic magnetic quantities, e.g. total magnetic and toroidal moments for the sinusoidally rippled film and a roll-up nanotube, respectively. Since the macroscopic quantities scale with the film volume, they
can be made large enough to be accessible experimentally.

\section*{Acknowledgments}
This work was supported by the Deutsche Forschungsgemeinschaft (DFG, German Research Foundation) through the Sonderforschungsbereich SFB 1143, grant No. YE 232/1-1, and under Germany's Excellence Strategy through the W\"{u}rzburg-Dresden Cluster of Excellence on Complexity and Topology in Quantum Matter -- \emph{ct.qmat} (EXC 2147, project-ids 390858490 and 392019). O.G. and J.S. acknowledge funding by the Deutsche Forschungsgemeinschaft (DFG, German Research Foundation)-TRR288-422213477 (project A09, and A12)  and TRR 173 – 268565370 (project A09, A11, and B15).

\appendix

\section{Discrete and continuous models}\label{app:models}
For the case of a planar film, Hamiltonian of the discrete model shown in Fig.~\ref{fig:bending}(b) can be written in the  form
\begin{subequations}\label{eq:H-discr}
    \begin{align}
\mathcal{H}=\mathcal{H}_{12}+\mathcal{H}_1+\mathcal{H}_2,
\end{align}
where the first summand represents the inter-sublattice antiferromagnetic exchange with $J>0$: 
\begin{align}
&\mathcal{H}_{12}=J\sum\limits_{\vec{R}_{\vec{n}}}\hspace{-1cm}\sum\limits_{\hspace{1cm}\sigma\in\{\nearrow,\swarrow,\nwarrow,\searrow\}}\hspace{-1cm}\vec{\mathcal{M}}_1(\vec{R}_{\vec{n}})\cdot\vec{\mathcal{M}}_2(\vec{R}_{\vec{n}}+\delta\vec{R}_{\sigma}).
    \end{align}
Here the unit vector $\vec{\mathcal{M}}_\nu$ with $\nu=1,2$ denotes normalized magnetic moment of the $\nu$-th sublattice, and the position vector $\vec{R}_{\vec{n}}=a_0(n_1\tilde{\vec{e}}_1+n_2\tilde{\vec{e}}_2)$ numerates nodes of the first sublattice. Additionally, we introduce the following shift vectors $\delta\vec{R}_{\nearrow}=-\delta\vec{R}_{\swarrow}=\frac{a_0}{2}(\tilde{\vec{e}}_1+\tilde{\vec{e}}_2)$, $\delta\vec{R}_{\nwarrow}=-\delta\vec{R}_{\searrow}=\frac{a_0}{2}(-\tilde{\vec{e}}_1+\tilde{\vec{e}}_2)$. Hamiltonians $\mathcal{H}_{1,2}$ represent the contributions from each of the sublattices, they are as follows
\begin{align}
    \mathcal{H}_1=&\sum\limits_{\vec{R}_{\vec{n}}}\biggl\{\frac{\tilde{J}}{2}\biggl[\hspace{-0.7cm}\sum\limits_{\hspace{0.7cm}\sigma\in\{\nwarrow,\searrow\}}\hspace{-0.7cm}\vec{\mathcal{M}}_1(\vec{R}_{\vec{n}})\cdot\vec{\mathcal{M}}_1(\vec{R}_{\vec{n}}+2\delta\vec{R}_{\sigma})\\\nonumber
    &-\hspace{-0.7cm}\sum\limits_{\hspace{0.7cm}\sigma\in\{\nearrow,\swarrow\}}\hspace{-0.7cm}\vec{\mathcal{M}}_1(\vec{R}_{\vec{n}})\cdot\vec{\mathcal{M}}_1(\vec{R}_{\vec{n}}+2\delta\vec{R}_{\sigma})\biggr]\\\nonumber
    &-\tilde{K}(\vec{\mathcal{M}}_1(\vec{R}_{\vec{n}})\cdot\vec{e}_3)^2-\mu_s\vec{\mathcal{M}}_1(\vec{R}_{\vec{n}})\cdot\vec{B}\biggr\},\\
    \mathcal{H}_2=&\sum\limits_{\vec{R}'_{\vec{n}}}\biggl\{-\frac{\tilde{J}}{2}\biggl[\hspace{-0.7cm}\sum\limits_{\hspace{0.7cm}\sigma\in\{\nwarrow,\searrow\}}\hspace{-0.7cm}\vec{\mathcal{M}}_2(\vec{R}'_{\vec{n}})\cdot\vec{\mathcal{M}}_2(\vec{R}'_{\vec{n}}+2\delta\vec{R}_{\sigma})\\\nonumber
    &-\hspace{-0.7cm}\sum\limits_{\hspace{0.7cm}\sigma\in\{\nearrow,\swarrow\}}\hspace{-0.7cm}\vec{\mathcal{M}}_2(\vec{R}'_{\vec{n}})\cdot\vec{\mathcal{M}}_2(\vec{R}'_{\vec{n}}+2\delta\vec{R}_{\sigma})\biggr]\\\nonumber
    &-\tilde{K}(\vec{\mathcal{M}}_2(\vec{R}'_{\vec{n}})\cdot\vec{e}_3)^2-\mu_s\vec{\mathcal{M}}_2(\vec{R}'_{\vec{n}})\cdot\vec{B}\biggl\},
\end{align}
\end{subequations}
where $\vec{R}'_{\vec{n}}=\vec{R}_{\vec{n}}+\delta\vec{R}_{\nearrow}$ numerates nodes of the second sublattice. Note the sign flip at the front of the altermagnetic terms. Hamiltonian \eqref{eq:H-discr} is a simplified version of the Hamiltonian previously proposed~\cite{Gomonay24} for RuO$_2$ in which we additionally introduce the interaction with external magnetic field.

To obtain the continuous approximation of Hamiltonian \eqref{eq:H-discr}, we utilize the Taylor expansion
\begin{equation}
\begin{split}
\vec{\mathcal{M}}_\nu&(\vec{R}_{\vec{n}}+\delta\vec{R}_\sigma)\approx\vec{\mathcal{M}}_\nu(\vec{R}_{\vec{n}})+(\delta\vec{R}_\sigma)_{\tilde\alpha}\partial_{\tilde\alpha}\vec{\mathcal{M}}_{\nu}(\vec{R}_{\vec{n}})\\
&+\frac12(\delta\vec{R}_\sigma)_{\tilde\alpha}(\delta\vec{R}_\sigma)_{\tilde\beta}\partial^2_{\tilde\alpha\tilde\beta}\vec{\mathcal{M}}_{\nu}(\vec{R}_{\vec{n}})
\end{split}
\end{equation}
with $\tilde\alpha,\tilde\beta\in\{\tilde\xi_1,\tilde\xi_2\}$ and we denote $\partial_{\tilde\alpha}=\partial_{\tilde\xi_\alpha}$. Next we replace the summation by integration: $\sum_{\vec{R}_{\vec{n}}}(\dots)\to\frac{1}{a_0^2c_0}\int(\dots)$ performing a primitive generalization to 3D case. This enables us to present hamiltonian \eqref{eq:H-discr} in the following form $\mathcal{H}=\int\mathscr{E}\dd\tilde{\xi}_1\dd\tilde{\xi}_2\dd\tilde\xi_3$ with
\begin{equation}\label{eq:H-cnt}
\begin{split}
&\mathscr{E}=\frac12B_{ex}M_s\,\vec{\mathcal{M}}_1\cdot\vec{\mathcal{M}}_2-A_{\textsc{afm}}\partial_{\tilde\alpha}\vec{\mathcal{M}}_1\cdot\partial_{\tilde\alpha}\vec{\mathcal{M}}_2\\
&+\frac{A_{\textsc{alt}}}{2}\left(\partial_{\tilde1}\vec{\mathcal{M}}_1\cdot\partial_{\tilde2}\vec{\mathcal{M}}_1-\partial_{\tilde1}\vec{\mathcal{M}}_2\cdot\partial_{\tilde2}\vec{\mathcal{M}}_2\right)\\
&-\sum\limits_{\nu=1,2}\left[K(\vec{e}_3\cdot\vec{\mathcal{M}}_{\nu})^2+M_s(\vec{B}\cdot\vec{\mathcal{M}}_{\nu})\right]
\end{split}
\end{equation} 
being the energy density.
Here $K=\tilde{K}/(a_0^2c_0)$ and the rest of the constants are defined in the main text.

Dynamics of the vector fields $\vec{\mathcal{M}}_{1,2}$ is governed by the set of two coupled Landau-Lifshitz equations
\begin{equation}\label{eq:LL-cnt}
\partial_t\vec{\mathcal{M}}_\nu=\frac{\gamma_0}{M_s}\left[\vec{\mathcal{M}}_\nu\times\frac{\delta\mathcal{H}}{\delta\vec{\mathcal{M}}_\nu}\right],\qquad \nu=1,\,2.
\end{equation}
Next, we introduce the dimensionless N{\'e}el vector $\vec{n}=(\vec{\mathcal{M}}_1-\vec{\mathcal{M}}_2)/2$ and the magnetization vector $\vec{m}=(\vec{\mathcal{M}}_1+\vec{\mathcal{M}}_2)/2$. Taking into account that $\vec{n}^2+\vec{m}^2=1$ and $\vec{n}\cdot\vec{m}=0$, we present energy density \eqref{eq:H-cnt} in form 
\begin{equation}\label{eq:H}
\begin{split}
\mathscr{E}&=B_{ex}M_s\vec{m}^2+A_{\textsc{afm}}\,\partial_{\tilde{\alpha}}\vec{n}\cdot\partial_{\tilde\alpha}\vec{n}-K(\vec{n}\cdot\vec{e}_3)^2\\
&+A_{\textsc{alt}}(\partial_{\tilde{1}}\vec{m}\cdot\partial_{\tilde{2}}\vec{n}+\partial_{\tilde{2}}\vec{m}\cdot\partial_{\tilde{1}}\vec{n})-2M_s\vec{B}\cdot\vec{m},
\end{split}
\end{equation}
where we neglected all terms quadratic in $m$ except the uniform exchange. In terms of $\vec{n}$ and $\vec{m}$, equations \eqref{eq:LL-cnt} obtain the following form 
\begin{equation}\label{eq:LL-nm}
\begin{split}
&\dot{\vec{n}}=\frac{\gamma_0}{2M_s}\left[\vec{m}\times\frac{\delta\mathcal{H}}{\delta\vec{n}}+\vec{n}\times\frac{\delta\mathcal{H}}{\delta\vec{m}}\right],\\
&\dot{\vec{m}}=\frac{\gamma_0}{2M_s}\left[\vec{n}\times\frac{\delta\mathcal{H}}{\delta\vec{n}}+\vec{m}\times\frac{\delta\mathcal{H}}{\delta\vec{m}}\right].
\end{split}
\end{equation}

In the first order in small parameter $|\vec{m}|\ll1$ and taking into account that $\sqrt{B_{an}/B_{ex}}\ll1$, we obtain from \eqref{eq:LL-nm}
\begin{align}\label{eq_dynamics_n}
\nonumber
&\bigl\{\ddot{\vec{n}}-c^2\nabla^2\vec{n}-\omega_{\textsc{afmr}}^2n_z\vec{e}_z+\Lambda\bigl(2\vec{n}\times\partial^2_{\tilde{1}\tilde{2}}\dot{\vec{n}}+\dot{\vec{n}}\times\partial^2_{\tilde{1}\tilde{2}}\vec{n}\\\nonumber
&+\partial_{\tilde{1}}\vec{n}\times\partial_{\tilde{2}}\dot{\vec{n}}+\partial_{\tilde{2}}\vec{n}\times\partial_{\tilde{1}}\dot{\vec{n}}\bigr)+\gamma_0\Lambda\bigl[(\vec{n}\cdot\vec{B})\partial_{\tilde{1}\tilde2}^2\vec{n}\\
&+\vec{B}(\vec{n}\cdot\partial_{\tilde{1}\tilde{2}}^2\vec{n})-\partial_{\tilde{1}\tilde{2}}^2(\vec{n}\times\vec{B}\times\vec{n})\bigr]\bigr\}\times\vec{n}=0.
\end{align}
Here $\omega_{\textsc{afmr}}=\gamma_0\sqrt{B_{ex}B_{an}}$ is frequency of the uniform ferromagnetic resonance, $c=\gamma_0\sqrt{B_{ex}A_{\textsc{afm}}/M_s}$ is maximal magnon velocity, and $\Lambda=\gamma_0 A_{\textsc{alt}}/M_s$ represents the altermagnetism strength. In this case, the magnetization is determined by formula \eqref{eq:m}. Eq.~\eqref{eq_dynamics_n} is the Euler-Lagrange equation for Lagrangian \eqref{eq:L} in which we performed the change of variables $\tilde{\xi}_\alpha\to\xi_\alpha$, see Fig.~\ref{fig:bending}(b).

After the bending shown in Fig.~\ref{fig:bending}(a), the chosen frame of reference $(\xi_1,\xi_2)$ preserves the Euclidean metric meaning that the Lagrangian \eqref{eq:L} retains its form also for the bent film. Note that this kind of generalization is not applicable for the surfaces with nonvanishing Gau{\ss}ian curvature.

\section{The effects of geometry}\label{app:geom}

Without magnetic field, energy density of a static solution is $\mathscr{E}_m=A_{\textsc{afm}}\,\partial_\alpha\vec{n}\cdot\partial_\alpha\vec{n}+K[\vec{n}\times\vec{e}_3]^2$. Taking into account the rules of the local basis differentiation $\partial_1\vec{e}_1=-\kappa\vec{e}_3$, $\partial_1\vec{e}_2=0$, $\partial_1\vec{e}_3=\kappa\vec{e}_1$, and $\partial_2\vec{e}_i=0$ which follows from Frenet-Serret formulas, we write the exchange contribution as follows 
\begin{equation}\label{eq:Eex}
\begin{split}
    \partial_\alpha\vec{n}\cdot\partial_\alpha\vec{n}=&(\partial_\alpha n_i)(\partial_\alpha n_i)+2\kappa(n_3\partial_1n_1-n_1\partial_1n_3)\\
    &+\kappa^2(n_1^2+n_3^2),
    \end{split}
\end{equation}
where $\alpha=1,2$ and $\vec{n}=n_i\vec{e}_i$ with $i=1,2,3$. In \eqref{eq:Eex}, the first summand represents the conventional exchange, while the second and third summands are the effective curvature-induced Dzyloshinskyi-Moriya and anisotropy interactions, respectively~\cite{Gaididei14}. For the angular parameterization $n_1=\sin\theta\cos\phi$, $n_2=\sin\theta\sin\phi$, and $n_3=\cos\theta$, we present the magnetic energy density as follows
\begin{equation}
\begin{split}
    \mathscr{E}_m=&A_{\textsc{afm}}\bigl\{(\partial_\alpha\theta)^2+\sin^2\theta\left[(\partial_\alpha\phi)^2-\kappa^2\sin^2\phi+\ell^{-2}\right]\\
    &+2\kappa(\cos\phi\partial_1\theta-\sin\theta\cos\theta\sin\phi\partial_1\phi)+\kappa^2\bigr\}.
    \end{split}
\end{equation}
The corresponding Euler-Lagrange equations 
\begin{subequations}
\begin{align}
\label{eq:theta-det}&\partial_{\alpha\alpha}^2\theta-\sin\theta\cos\theta\left[(\partial_\alpha\phi)^2-\kappa^2\sin^2\phi+\ell^{-2}\right]\\\nonumber
&-2\sin^2\theta\sin\phi\,\partial_1\phi=-\cos\phi\,\partial_1\kappa,\\
\label{eq:phi-det}&\partial_\alpha(\sin^2\theta\,\partial_\alpha\phi)+\kappa\sin^2\theta\sin\phi(2\partial_1\theta+\kappa\cos\phi)\\\nonumber
&=\sin\theta\cos\theta\sin\phi\,\partial_1\kappa
\end{align}
\end{subequations}
have solution $\theta=\theta(\xi_1)$, $\phi=0$ which turns Eq.~\eqref{eq:phi-det} to identity and transforms Eq.~\eqref{eq:theta-det} to Eq.~\eqref{eq:theta}. 

Since $\vec{n}$ depends only on coordinate $\xi_1$, we write $\hat{\mathfrak{D}}\vec{n}=-\sin\varphi_0\cos\varphi_0\partial_{11}^2\vec{n}$. Taking into account the rules of the local basis differentiation, one obtains $\vec{n}\times\partial_{11}^2\vec{n}\times\vec{n}=(\partial_{11}^2\theta+\partial_1\kappa)(\vec{e}_1\cos\theta-\vec{e}_3\sin\theta)=\ell^{-2}\sin\theta\cos\theta(\vec{e}_1\cos\theta-\vec{e}_3\sin\theta)$ where we utilize Eq.~\eqref{eq:theta} on the last step. Finally, we present magnetization \eqref{eq:m} in form $\vec{m}=-\mu_{\textsc{alt}}\sin(2\varphi_0)\sin\theta\cos\theta(\vec{e}_1\cos\theta-\vec{e}_3\sin\theta)$. For a particular case $\varphi_0=-\pi/4$, it coincides with \eqref{eq:m-smpl-theta}. The influence of the angle $\varphi_0$ on magnetization is demonstrated in Fig.~\ref{fig:m-q}.

\section{Spin-lattice simulations}\label{app:simuls}
The dynamics of magnetic moments are described by the Landau--Lifshits equations
\begin{equation}\label{eq:sim_LL}
    \left(1+\alpha^2_\textsc{g}\right)\frac{\mathrm{d}\vec{\mathcal{M}}_{\nu}}{\mathrm{d}t}=\frac{\gamma}{M_s}\left(1+\alpha_\textsc{g}\vec{\mathcal{M}}_{\nu}\times\right)\vec{\mathcal{M}}_{\nu}\times\frac{\partial\mathcal{H}}{\partial\vec{\mathcal{M}}_{\nu}},
\end{equation} 
where $\alpha_\textsc{g}$ is the Gilbert damping parameter and $\mathcal{H}$ is defined in~\eqref{eq:H-discr}. The dynamical problem is considered as a set of $3N_1 N_2$ ordinary differential equations~\eqref{eq:sim_LL} with respect to $3N_1 N_2$ unknown functions $\mathcal{M}_\nu^\textsc{x}(t),\ \mathcal{M}_\nu^\textsc{y}(t),\ \mathcal{M}_\nu^\textsc{z}(t)$. Parameters $N_1$ and $N_2$ define the size of the system. For the given geometry and initial conditions, the set of time evolution equations~\eqref{eq:sim_LL} is integrated numerically using the Runge–-Kutta method in Python.

In all simulations we use the following material parameters: Gilbert damping $\alpha_\textsc{g} = 0.5$, magnetic length $\ell = 10 a_0$, exchange stiffness ratio $A_\textsc{alt}/A_\textsc{afm} = 0.5$, and anisotropy/exchange fields ratio $B_{an}/B_{ex} = 6.25\times 10^{-4}$.

\subsection{Simulations of sinusoidal-shaped film}\label{sec:sim_sin}

We considered the sinusoidal surface with directrix $\vec{\gamma}=x\hat{\vec{x}}+f(x)\hat{\vec{y}}$ and $f=\mathscr{A}\cos(2\pi x/T)$. In simulations, we considered a sinusoidal surface with $N_1 = 1001$ and $N_2 = 101$, amplitudes $\mathscr{A} \in \left[2.5;7.5\right]\ell$, period $T \in \left[5;50\right]\ell$. These geometrical parameters result in films with more then 2 periods. 

The simulations were performed in one step. By setting an initial state as a normal state with $\vec{\mathcal{M}}_{\nu}||\vec{e}_3$ we run simulations for given geometry in a long time regime with $\Delta t \gg 1/\left(\alpha_\textsc{g} \omega_\textsc{afmr}\right)$. The resulting curvilinear components of magnetization vectors were defined as $m_i = \vec{m}\cdot\vec{e}_i$ and presented in Fig.~\ref{fig:m-sin}.

\subsection{Simulations of rolled up nanotube}\label{sec:sim_aSpiral}

We considered the rolled up nanotube with a shape of  cylindrical Archimedean spiral with directrix $\vec{\gamma}=r(\chi)(\hat{\vec{x}}\cos\chi+\mathcal{C}\hat{\vec{y}}\sin\chi)$ where $r=\delta_r\chi/(2\pi)$ and $\chi>0$. In simulations, we considered a sinusoidal surface with $N_1 = 1001$ and $N_2 = 101$, spiral step $\delta_r \in \left[0.5;2.5\right]\ell$.

Simulations for the rolled up nanotube were performed in the same manner as in Sec.~\ref{sec:sim_sin}. The resulting tangential component of magnetization $m_1 = \vec{m}\cdot\vec{e}_1$ is presented in Fig.~\ref{fig:m-aSpiral}.

\section{Self-consistent problem in applied magnetic field}\label{app:elast}

Here we discuss equilibrium solutions in the applied magnetic field when both magnetic and elastic energies are taken into account. In the following, we consider possible stretching-free deformations in form of a generalized cylindrical surface, see Fig.~\ref{fig:bending}. In this case, the elastic energy of a thin film can be approximated by the bending contribution only:~\cite{Efrati09,Armon11}
\begin{equation}\label{eq:E-el}
\begin{split}
    &E_{el}\approx\frac{Yh^3}{24(1+\nu)}\int\mathscr{E}_{\textsc{b}}\sqrt{\bar{g}}\,\dd\xi_1\dd\xi_2,\\    &\mathscr{E}_\textsc{b}=\left(\frac{\nu}{1-\nu}\bar{g}^{\alpha\beta}\bar{g}^{\gamma\delta}+\bar{g}^{\alpha\gamma}\bar{g}^{\beta\delta}\right)b_{\alpha\beta}b_{\gamma\delta},
\end{split}    
\end{equation}
were $Y$ is the Young's modulus,  $\nu$ is the Poisson's ratio, and $h$ is the film thickness.
Here $g_{\alpha\beta}=\partial_\alpha\vec{\sigma}\cdot\partial_\beta\vec{\sigma}$ with $\alpha,\beta\in\{\xi_1,\xi_2\}$ is the metric tensor and $\bar{g}_{\alpha\beta}$ denotes the reference metric of a film free of elastic tensions and $\bar{g}=\det[\bar{g}_{\alpha\beta}]$. As a reference state, we consider a planar film laying in $xz$-plane. Since $\xi_1$ is the arclength of the generatrix $\vec{\gamma}$, one has $g_{\alpha\beta}=\bar{g}_{\alpha\beta}=\delta_{\alpha\beta}$ due to the absence of the stretching. The bending contribution is determined by the elements of the second fundamental form $b_{\alpha\beta}=\vec{e}_3\cdot\partial_{\alpha\beta}^2\vec{\sigma}$ which for the geometry determined in Fig.~\ref{fig:bending} are $b_{11}=-\kappa$, $b_{22}=b_{12}=b_{21}=0$. Finally, $\mathscr{E}_\textsc{b}=\kappa^2/(1-\nu)$ and elastic energy \eqref{eq:E-el} is reduced to
\begin{equation}
    E_{el}=\frac{Yh^3}{24(1-\nu^2)}\int\kappa^2\dd\xi_1\dd\xi_2.
\end{equation}


Let us now consider magnetic energy of the deformed film in the applied magnetic field. From Lagrangian \eqref{eq:L} one obtains the energy density as a corresponding element of the energy-momentum tensor. For a static case, the magnetic energy density is
\begin{equation}\label{eq:E}
\begin{split}
    \mathscr{E}_m=&A_{\textsc{afm}}\,\partial_\alpha\vec{n}\cdot\partial_\alpha\vec{n}+K[\vec{n}\times\vec{e}_3]^2-\frac{M_s}{B_{ex}}[\vec{n}\times\vec{B}]^2\\
    &-\frac{2A_{\textsc{alt}}}{B_{ex}}(\vec{n}\times\vec{B}\times\vec{n})\cdot\hat{\mathfrak{D}}\vec{n}.
    \end{split}
\end{equation}
For the case $\vec{B}=B\hat{\vec{y}}$, and assuming that $\vec{n}=\vec{e}_1\sin\theta+\vec{e}_3\cos\theta$ with $\theta=\theta(\xi_1)$ we write \eqref{eq:E} in form
\begin{equation}
    \begin{split}
    \mathscr{E}_m=&K\bigl[\ell^2(\theta'+\kappa)^2+\sin^2\theta+(b_1\sin\theta+b_3\cos\theta)^2\\
    &+\varepsilon\ell^2\sin2\varphi_0(\theta''+\kappa')(b_1\cos\theta-b_3\sin\theta)\bigr],
    \end{split}
\end{equation}
where $\varepsilon=(A_{\textsc{alt}}/A_{\textsc{afm}})\sqrt{B_{an}/B_{ex}}$ represents strength of the altermagnetism and $\vec{b}=\vec{B}/\sqrt{B_{ex}B_{an}}$ is the magnetic field in units of the spin-flop field. As previously, prime denotes the derivative with respect to the arclength $\xi_1$. The film shape in encoded in the curvature $\kappa$ as well as in the magnetic field components $b_1=\vec{b}\cdot\vec{e}_1$ and $b_3=\vec{b}\cdot\vec{e}_3$. In the following, it is instructive to describe the geometrical film profile in terms of the angle $\psi=\angle(\vec{e}_1,\hat{\vec{x}})$ of inclination of the film. In the other words, $\tan\psi=f'_x$. In this case, $\kappa=-\psi'$, $b_1=b\sin\psi$, and $b_3=b\cos\psi$. Now, the total energy functional can be presented as follows
\begin{equation}\label{eq:Etot}
\begin{split}
    E_{\text{tot}}&=Kh\int\bigl\{\ell^2[1+\varepsilon b\sin2\varphi_0\cos(\theta-\psi)](\theta'-\psi')^2\\
    &+\sin^2\theta+b^2\cos^2(\theta-\psi)+Q^{-2}\psi'^2\bigr\}\dd\xi_1\dd\xi_2,
    \end{split}
\end{equation}
where the last summand with $Q=h^{-1}\sqrt{24K(1-\nu^2)/Y}$ repersents the bending energy. For the case $\varepsilon b<1$ (a realistic situation), the energy \eqref{eq:Etot} is minimized for $\theta=0$ and $\psi=\pm\pi/2$. This solution realizes a geometrical spin-flop, when the film is oriented parallel to the field, and therefore, the N{\'e}el order parameter is perpendicular to the field. Note that $b^2$-term is responsible for the geometric spin-flop. In the following, we consider the case of small fields $B\ll B_{an}(A_{\textsc{alt}}/A_{\textsc{afm}})$, which allows us to limit ourselves with only the linear in $b$ term which is of purely altermagnetic nature. Our aim is to analyze stability of the planar plane solution ($\theta=0$, $\psi=0$) with respect to the applied perpendicular magnetic field. For small $\theta$ and $\psi$, we simplify \eqref{eq:Etot} to
\begin{equation}\label{eq:Rtot-smpl}
    E_{\text{tot}}\approx Kh\int\left[k^{-2}(\theta'-\psi')^2+\theta^2+Q^{-2}\psi'^2\right]\dd\xi_1\dd\xi_2,
\end{equation}
where $k=1/(\ell\sqrt{1+b\varepsilon\sin2\varphi_0})$. From \eqref{eq:Rtot-smpl}, we obtain the relation $\psi'=\theta'Q^2/(k^2+Q^2)$ and exclude $\psi'$:
\begin{equation}\label{eq:Rtot-smpl-smpl}
    E_{\text{tot}}\approx Kh\int\left[\frac{\theta'^2}{Q^2+k^2}+\theta^2\right]\dd\xi_1\dd\xi_2,
\end{equation}
The solution $\theta=0$ is stable for $Q^2+k^2>0$. The latter condition is fulfilled for $\varepsilon b<1$. The corresponding solution $\psi'=0$ means that the film is flat, however, it can be arbitrary inclined to the field.



%

\end{document}